\documentclass[twocolumn]{aastex631} %, linenumbers

\usepackage{graphics,epsf}
\usepackage[utf8]{inputenc}
\usepackage{amsmath}                % American Mathematical Society package
\usepackage{amsfonts}               % American Mathematical Society fonts
\usepackage{amssymb}                % American Mathematical Society symbol
\usepackage{epsfig}                 % EPS figures
\usepackage{graphicx}               % Required for inserting images
\usepackage{float}
\usepackage{color}
\usepackage{multirow}               % double row table entries
\usepackage{hyperref}

\hypersetup{
    colorlinks=true,
    linkcolor=red,   
    urlcolor=cyan}

\usepackage[colorinlistoftodos]{todonotes}

\newcommand{\cm}{{~\rm cm}}
\newcommand{\km}{{~\rm km}}
\newcommand{\s}{{~\rm s}}

\newcommand{\g}{{~\rm g}}
\newcommand{\G}{{~\rm G}}
\newcommand{\K}{{~\rm K}}
\newcommand{\erg}{{~\rm erg}}

\begin{document}

\title{The two alternative explosion mechanisms of core-collapse supernovae:  2024 status report}
\date{October 2024}

\author[0000-0003-0375-8987]{Noam Soker}
\affiliation{Department of Physics, Technion, Haifa, 3200003, Israel; soker@physics.technion.ac.il}

\begin{abstract}
In comparing the two alternative explosion mechanisms of core-collapse supernovae (CCSNe), I examine recent three-dimensional (3D) hydrodynamical simulations of CCSNe in the frame of the delayed-neutrino explosion mechanism (neutrino mechanism) and argue that these valuable simulations show that neutrino heating can supply a non-negligible fraction of the explosion energy but not the observed energies, hence cannot be the primary explosion mechanism. In addition to the energy crisis, the neutrino mechanism predicts many failed supernovae that are not observed. The most challenging issue of the neutrino mechanism is that it cannot account for point-symmetric morphologies of CCSN remnants, many of which were identified in 2024. These contradictions with observations imply that the neutrino mechanism cannot be the primary explosion mechanism of CCSNe. The alternative jittering-jets explosion mechanism (JJEM) seems to be the primary explosion mechanism of CCSNe; neutrino heating boosts the energy of the jittering jets. Even if some simulations show explosions of stellar models (but usually with energies below observed), it does not mean that the neutrino mechanism is the explosion mechanism. Jittering jets, which simulations do not include, can explode the core before the neutrino heating process does. Morphological signatures of jets in many CCSN remnants suggest that jittering jets are the primary driving mechanism, as expected by the JJEM.    

\end{abstract}


% ================================================
\section{Introduction}
\label{sec:Introduction}
% ================================================

Two theoretical alternative explosion mechanisms that are based on the gravitational energy of the collapsing stellar core exist to explode core-collapse supernovae (CCSNe): the delayed neutrino explosion mechanism (neutrino mechanism) and the jittering jets explosion mechanism (JJEM). In 2024, more than a dozen papers on each mechanism further explored their properties. 

Studies of the neutrino mechanism in 2024 have concentrated mainly on simulating the outcomes of neutrino heating of the collapsing core of the stellar progenitor   (e.g., \citealt{Andresenetal2024, Burrowsetal2024kick, JankaKresse2024, Muler2024, Mulleretal2024, Nakamuraetal2024, vanBaaletal2024, WangBurrows2024, Laplaceetal2024, Huangetal2024, Bocciolietal2025, Maunderetal2025}). Neutrino heating occurs behind the stalled shock (closer to the newly born neutron star; NS) in the gain region at radii of $r \simeq 100 \km$. Jets play no role in these simulations. I group the jet-based magnetorotational explosion mechanism (e.g., \citealt{Kondratyevetal2024, Shibagakietal2024, ZhaMullerPowell2024}) with the neutrino mechanism despite that it is based on jets because this mechanism operates only in rare cases ($\simeq 1\%$ of CCSNe, e.g., \citealt{Muller2024}) when there is a rapidly rotating pre-collapse core; this mechanism assumes that most CCSNe are jetless and explode by the neutrino mechanism. 

In the JJEM, the newly born NS, or later a black hole if formed, launches several to tens of pairs of opposite jets with fully or partially stochastically varying directions; these jets explode the star. In \citep{Soker2024Learning}, I list the updated quantitative estimates of the properties of the jittering jets for most CCSNe that are the descendant of iron core collapse to an NS. The properties of the JJEM in electron capture CCSNe (if they occur) were studied by \cite{WangShishkinSoker2024}.

Many expected outcomes of the JJEM are similar (but not identical) to those of the neutrino mechanism, like overall nucleosynthesis and neutrino emission; some other emission differences are hard to detect, like different gravitational wave properties (e.g., \citealt{Mezzacappaetal2023} for the neutrino mechanism and \citealt{Soker2023GW} for the JJEM). The most robust property to distinguish the predictions of the neutrino mechanism and the JJEM is the morphology of CCSN remnants (CCSNRs), specifically point-symmetric morphologies. In Section \ref{sec:comparison}, I elaborate on these common properties and differences following 2024 papers. 

The JJEM predicts that in many, but not all, CCSNRs, two or more pairs of jittering jets will imprint opposite (to the center) structural features. Namely, the jittering jets will shape a point-symmetric CCSNR.
At the time of submission of this paper, the neutrino mechanism does not explain most of the properties of point-symmetric CCSNRs, and it seems it cannot explain point-symmetric CCSNRs (\citealt{SokerShishkin2024Vela}). 
Indeed, in a recent study \cite{Vartanyanetal2025} present a simulation of the neutrino mechanism until shock breakout from the star. They do not reproduce a point-symmetric morphology and do not refer to it.  
For these, I consider point-symmetric CCSNRs to rule out the neutrino mechanism as the primary explosion mechanism of CCSNRs. 

The systematic identification of point-symmetric CCSNRs and their attribution to the JJEM was a breakthrough in 2023 (\citealt{Soker2024Rev} for a review), with more CCSNRs identified in 2024. 
The list of 12 point-symmetric CCSNRs with attributed morphologies to the JJEM is as follows: 
SNR 0540-69.3 \citep{Soker2022SNR0540},
the Vela CCSNR (\citealt{Soker2023SNRclass, SokerShishkin2024Vela}), 
CTB~1 \citep{BearSoker2023RNAAS}, 
Cassiopeia A \citep{BearSoker2024}, 
Puppis A \citep{Bearetal2024Puppis},
the Cygnus Loop \citep{ShishkinKayeSoker2024},
N63A \citep{Soker2024CounterJet}, 
SN 1987A \citep{Soker2024NA1987A, Soker2024Keyhole}, 
G321.3–3.9 \citep{Soker2024CF, ShishkinSoker2025G321}, 
G107.7-5.1 \citep{Soker2024CF}, 
W44 \citep{Soker2025W44}, 
and the Crab Nebula \citep{ShishkinSoker2025Crab}.

In a review talk at the Transients Down Under meeting\footnote{\url{https://transientsdownunder.github.io/program_slides/}}, held in Melbourne, Australia, on 29 January 2024, Hans-Thomas Janka reported on the status of neutrino-driven explosions in CCSN simulations and writes about the simulations of different research groups: ``3D simulations differ in many aspects of numerics, physics inputs, seed perturbations, and, qualitatively and quantitatively, in their outcomes.
3D code comparison is missing and highly desirable.'' 
Indeed, different groups that study the neutrino mechanism obtain different results, not only in three-dimensional (3D) simulations but also in scaled 1D simulations, such as which initial stellar mass ends in an explosion or `failed supernova' (e.g., figure 1 in \citealt{BoccioliFragione2024PRD} and Figures 5 and 6 in \citealt{BoccioliRoberti2024Univ} comparing to \citealt{Sukhboldetal2016}). 
A comparison between these groups of simulators is indeed highly desirable. I also find that comparing the neutrino mechanism and the JJEM is highly desirable. Therefore, I set the goal of performing such a comparison in this paper (Section \ref{sec:comparison}). In this paper, I focus on new results from 2024 (for a partial comparison of earlier studies of the two alternative CCSN explosion mechanisms, see the review \citealt{Soker2024Rev}). 

In Section \ref{sec:Energy}, I examine two recent sets of simulations of the neutrino mechanism to emphasize the point that that mechanism comes short of supplying the observed explosion energies. 
In Section \ref{sec:Convection} I discuss the pre-collapse core convection and its role in the two mechanisms, possibly also in pre-explosion envelope activity. 
In Section \ref{sec:Summary}, I summarize this comparison and conclude that even if simulations of the neutrino mechanism manage to explode some stellar models, it does not imply that it is the primary explosion mechanism of CCSNe.

% ================================================
\section{Comparing the two mechanisms}
\label{sec:comparison}
% ================================================

In this section, I compare several outcomes of the two alternative explosion mechanisms and one requirement for an explosion common to both models. I elaborate on the last two in sections \ref{sec:Energy} and \ref{sec:Convection}, respectively.    

\begin{enumerate}
    \item \textit{Neutrino emission.} The neutrino emission was studied in many papers in the frame of the neutrino mechanism (e.g., \citealt{Fiorilloetal2023}). As most of the neutrinos result from the cooling of the NS, the neutrino emission expected in the JJEM is the same as that of the neutrino mechanism. Neutrino emission cannot distinguish between the two mechanisms, and in any case, was observed only from SN 1987A.  
    \item \textit{Gravitaional waves.} In the neutrino mechanism, convection behind the stalled shock and in the NS and the standing accretion shock instability emit gravitational waves in the crude frequency range of $100–2000 ~{\rm Hz}$ (e.g., \citealt{Mezzacappaetal2023}). 
    In the JJEM, the source of the gravitational waves is assumed to be the turbulent bubbles (cocoons) that the jittering jets inflate by interacting with the outer layers of the stellar core thousands of kilometers from the NS; the gravitational wave frequency is $\simeq 10-30 ~{\rm Hz}$ \citep{Soker2023GW}. The process of gravitational wave emission in the JJEM requires further study. The gravitational wave detectors are not yet in the stage to distinguish between the two mechanisms. 
    \item \textit{Nucleosynthesis.} There are no simulations of the nucleosynthesis in the JJEM. However, if the explosion energy and duration are about the same for a given CCSN, the expected nucleosynthesis in the JJEM is similar, but not identical, to that of the neutrino mechanism. However, some isotopes' geometrical distribution might differ in the two models. Particularly, the JJEM accounts for an `S-shaped' distribution of ejecta material as the observed O/Ne/Mg morphology in the Vela CCSNR (\citealt{SokerShishkin2024Vela}; for the observation, see \citealt{Mayeretal2023}).
    \item \textit{Lightcurve and spectrum.} For a given progenitor and explosion energy, the light curves and spectrum of the CCSN should be similar according to the neutrino mechanism and the JJEM. Even the polarization might not distinguish between the two; the JJEM has jets that can form elongated structures, but so are instabilities in the neutrino mechanism. In cases where jets break out, the JJEM might predict high-velocity components of ejecta, but it is unclear how strong their signatures are on the spectrum.      
    \item \textit{Black hole formation.} One prediction of the neutrino mechanism is that some stars do not explode but rather collapse to form a black hole in a `failed supernova;' there might be a faint transient event. The JJEM predicts no failed CCSNe. One `failed- supernova' candidate was N6946-BH1 (e.g., \citealt{Adamsetal2017}). An alternative explanation was immediately proposed, that of a type II intermediate luminosity optical transient (ILOT), where the merger of two stars expels equatorial ejecta that obscures the merger remnant, or the binary stellar system \citep{KashiSoker2017, SokerTypeII2021, BearetalTypeII2022}. \cite{Beasoretal2024} detected a luminous infrared source at the position of N6946-BH1, implying, as they conclude, that it is not a `failed supernova' but most likely a type II ILOT (for a different view, see \citealt{Kochanek2024, Kochaneketal2024}). More generally, recent studies question the existence of a population of `failed supernovae' \citep{ByrneFraser2022, StrotjohannOfekGalYam2024, BeasoretalLuminosity2024}.  The most recent study by \cite{BeasoretalLuminosity2024} deduces that the luminosity inferred for many CCSN progenitors is underestimated. This implies that many progenitors have masses $ > 20 M_\odot$, reducing the possible fraction of failed CCSN to be very low (and I argue, zero). 
    
    In a very recent paper, \cite{Deetal2024} claim that the faded star M31-2014-DS1 was a `failed supernova.' I find their arguments weak and argue it is also a type II ILOT. (a) They study the dust properties in a spherical geometry. Type II ILOT has the dust in an expanding equatorial outflow. Such an analysis, as with N6946-BH1 \citep{KashiSoker2017, SokerTypeII2021}, can also show the feasibility of the type II ILOT scenario for M31-2014-DS1. With their spherical analysis, \cite{Deetal2024} take N6946-BH1 to be a `failed supernova,' which is not \citep{Beasoretal2024}. (b) Their progenitor model is of a star with an initial mass of $M_{\rm ZAMS} \simeq 20 M_\odot$ and with a mass of only $M \simeq 6.7 M_\odot$ and a radius of $\simeq 400 R_\odot$ at explosion. To lose two-thirds of its mass, such a star likely had a binary interaction, as a single star of $M_{\rm ZAMS} \simeq 20 M_\odot$ is not expected to lose such a large mass (e.g., \citealt{Zapartasetal2025}). Such an interaction spun up the progenitor. Even if it were spun up to only $\simeq 0.001$ of its break-up velocity, i.e., a specific angular momentum of $j\simeq 10^{17} \cm^2 \s^{-1}$ of the outer envelope layer, the outer layers of the progenitor would form an accretion disk around the black hole. Such a disk would launch energetic jets. 
    (c) Even if the progenitor were rotating below $0.001$ of its break-up velocity, the envelope convection would form intermittent accretion disks around the black hole (e.g., \citealt{Quataertetal2019, AntoniQuataert2022, AntoniQuataert2023}). Indeed, \cite{Deetal2024} noticed that the envelope outer layer of $\simeq 0.15 M_\odot$ has the properties to form intermittent accretion disks. the ejected mass is $\simeq 0.1 M_\odot$. Even if only $0.01 M_\odot$ is accreted through intermittent accretion disks and launches jets, namely, the JJEM, and releases $5 \%$ of its mass, the explosion energy is $0.0005 M_\odot c^2 \simeq 10^{51} \erg$; There is an explosion in the frame of the JJEM.   {I predict the reappearance of M31-2014-DS1 in several years. }

   {I conclude from the discussion above that the neutrino mechanism's prediction of many failed CCSNe contradicts recent observations. }

   \item \textit{NS natal kick velocity.} The NS kick velocity in the neutrino mechanism is mainly due to asymmetrical mass ejection (e.g., as in Cassiopeia A, e.g., \citealt{HwangLaming2012}); the tugboat process (e.g., \citealt{Schecketal2004,  Nordhausetal2010, Wongwathanaratetal2013kick, Janka2017}). There are suggestions for other natal kick processes, e.g., \citealt{YamasakiFoglizzo2008, Yaoetal2021, Xuetal2022}; see \citealt{Igoshev2020} for general discussion). These processes, particularly the tugboat, can also operate in the JJEM. In addition, \cite{Bearetal2024Puppis} recently proposed the kick-BEAP (kick by early asymmetrical pair) to operate in the JJEM. In the kick-BEAP process, the NS launches a powerful pair of jets very early, within $\simeq 0.2 \s$ of bounce. One jet is much more powerful than the other, imparting a kick to the very young NS. The tugboat mechanism can also operate later in the explosion process. The kick-BEAP explains why some NSs have kick-spin alignment (if later mass-accretion episodes do not change the NS spin much) and some NSs have kick-spin misalignment (if later mass-accretion episodes change the NS spin). The kick-BEAP and tugboat processes that operate in the JJEM also account for the tendency of the angle between the main jet axis in CCSNRs and the kick direction to avoid small angles to each other, as observed (e.g., \citealt{BearSoker2018kick, BearSoker2023RNAAS}).  
    \item \textit{Morphologies of CCSNRs.} The JJEM expects some of the $\approx 5-30$ pairs of jets that explode the star to leave morphological imprints on the descendant CCSNR; instabilities in the explosion process and interaction with the interstellar medium and circumstellar material might smear these features in some CCSNRs. Although the neutrino mechanism might account for one opposite pair of morphological features by post-explosion pair of jets (e.g., \citealt{Orlandoetal2021} for Cassiopeia A), it cannot, in general, explain the properties of point-symmetric CCSNRs (\citealt{SokerShishkin2024Vela}; see also Section \ref{sec:Introduction} here). { The robust point-symmetric morphologies of some of the 12 CCSNRs with point symmetric morphologies decisively strongly support the JJEM, pose a severe challenge to the neutrino-driven explosion mechanism, and even rule it out as the primary explosion mechanism.} These CCSNRs with robust point-symmetric morphologies include SNR 0540-69.3 \citep{Soker2022SNR0540}, the Vela CCSNR \citep{SokerShishkin2024Vela}; Cassiopeia A \citep{BearSoker2024}; N63A \citep{Soker2024CounterJet}; and the Crab Nebula \citep{ShishkinSoker2025Crab}.    
    \item \textit{Simulated explosion energies in the neutrino mechanism.} 
    The most significant advantage of the neutrino mechanism over the JJEM is that {simulations of the delayed neutrino explosion mechanism in 3D have managed to explode some stellar models.} 
    On the other hand, the results of the simulations revealed problems with the neutrino mechanism, including disagreement between different groups that simulate the neutrino mechanism, the prediction of failed CCSNe (point 5 above), and the too-low explosion energies, which I elaborate on more in Section \ref{sec:Energy}.  
    There are no full simulations of the JJEM, as the required numerical grid resolution is several times smaller than the resolution of present CCSN simulations \citep{Soker2024Learning}.  
   \item \textit{Pre-collapse seed perturbations.} Both the neutrino mechanisms and the JJEM require large pre-collapse velocity and/or density perturbations (seed perturbations) in the core somewhere in the mass coordinate of $m \simeq 1.2-2 M_\odot$. The pre-collapse core convection in the silicon and oxygen-burning shells forms these seed perturbations. The neutrino mechanism requires the perturbation to form a non-smooth (non-spherical) surface of the stalled shock, the shock of the collapsing core at $r \simeq 150 \km$ from the NS, and to amplify the turbulence in the gain region behind the stalled shock; these facilitate shock revival (e.g., \citealt{Couchetal2015, Mulleretal2017}; Janka's 2024 talk cited in Section \ref{sec:Introduction}).  The JJEM requires the perturbation to be seed angular momentum perturbations, which, after amplified by instabilities behind the stalled shock, lead to the formation of the intermittent accretion disks that launch the jittering jets (e.g.,  \citealt{GilkisSoker2014}).   
   For the importance of the pre-collapse seed perturbation, I further explore the relation between the two models, and possibly with pre-explosion stellar activity, in Section \ref{sec:Convection}. 
\end{enumerate}

I turn to elaborate on points 8 (Section \ref{sec:Energy}) and 9 (Section \ref{sec:Convection}). 

% =========================================
\section{The explosion energy and NS mass}
\label{sec:Energy}
% =========================================

In \cite{Soker2024ggi}, I critically examined the 3D simulations by \cite{Nakamuraetal2024} and their claim that they find that neutrino-driven explosions occur for all models within 0.3~s after bounce (the time of shock formation at the center when collapse halts at about nuclear densities). \cite{Nakamuraetal2024} did not consider stellar material's binding energy (overburden) above the expanding shock. When I included the binding energy, I found the explosion energies they obtain to be $E_{\rm exp} \lesssim 0.1 \times 10^{51} \erg$ in some simulations and negative energy in some other simulations, implying they do not explode. 

\begin{figure*}[t]
\begin{center}
\includegraphics[trim=1cm 15.5cm 4.0cm 0.0cm,width=0.9\textwidth]{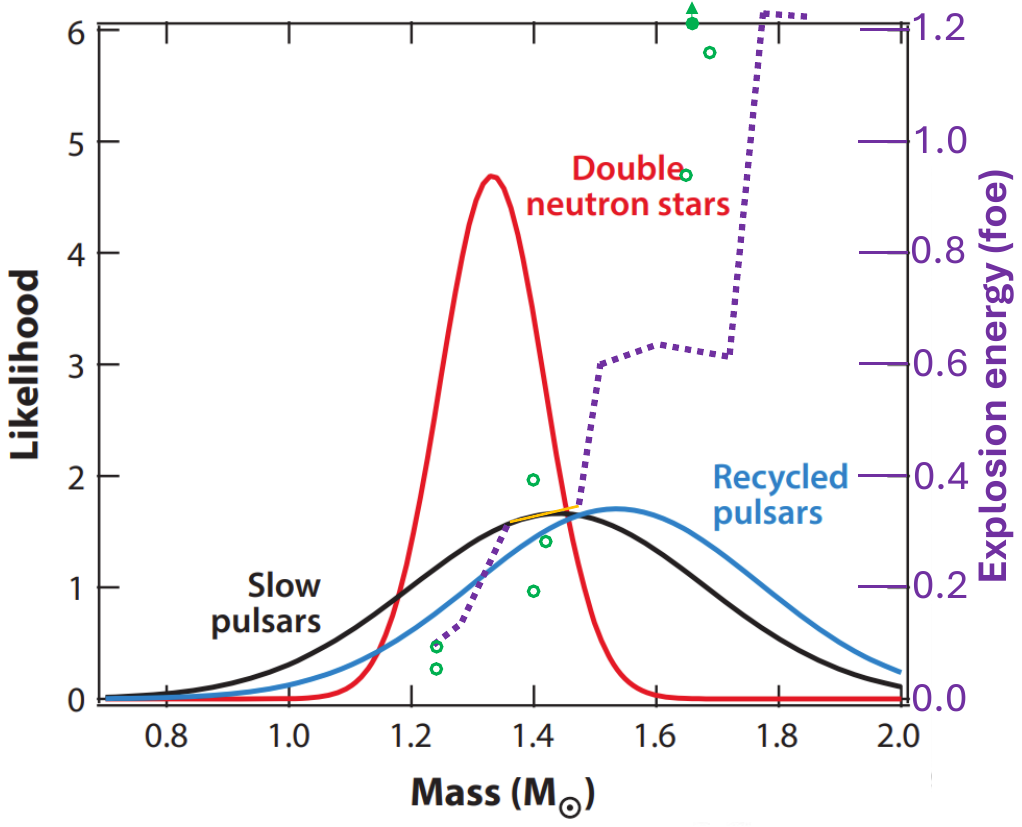} %[trim=left lower right upper]
\caption{Mass distribution of NSs adapted from \cite{OzelFreire2016} (three solid lines; scale on left). With a dashed-purple line and one solid-orange line, I added the explosion energy as a function of NS mass as \cite{Burrowsetal2024} obtained in their simulations of the neutrino mechanism (see text for more details). The units of explosion energy (right axis) are ${\rm foe}=10^{51} \erg$.  
The only simulations of \cite{Burrowsetal2024} that yield explosion energies of $E_{\rm exp} > 0.65 \times 10^{51} \erg$ are those that end with $M_{\rm NS} > 1.7 M_\odot$. 
With an open-green circle and one filled, I place the eight simulated cases by \cite{JankaKresse2024}; the filled circle is for a simulation with an explosion energy of $E_{\rm exp} = 1.46 \times 10^{51} \erg$. The results of \cite{JankaKresse2024} are similar to those of \cite{Burrowsetal2024}. These simulations yield $E_{\rm exp} > 0.5 \times 10^{51} \erg$ for $M_{\rm NS} \gtrsim 1.5 M_\odot$, which amounts to less than half of all NSs, contradicting the inferred explosion energy distribution of  CCSNe (see text). 
}
\label{fig:Expenergy}
\end{center}
\end{figure*}

In another 2024 paper, \cite{Burrowsetal2024} simulated the evolution of stellar models with initial masses of $M_{\rm ZAMS}=9-60 M_\odot$. Instead of examining the initial masses against observations, I examine the final NS masses. I take the observed NS mass distribution from \cite{OzelFreire2016}; the newer distribution by \cite{Meskhietal2022} is very similar. I adapt a figure from \cite{OzelFreire2016} and present it as the three thick solid lines, black, blue, and red, in Figure \ref{fig:Expenergy}. 
% FFFFFFFFFFFFFFFFFFFFFFFFFFFFFFFFFFFFFFFFFF
% FFFFFFFFFFFFFFFFFFFFFFFFFFFFFFFFFFFFFFFFFF

For each stellar model, \cite{Burrowsetal2024} give the remnant's mass and the explosion energy; I draw on Figure \ref{fig:Expenergy} with a dotted-purple line the energy (scale on the right axis in foe$=10^{51} \erg$) as a function of the gravitational NS mass (which is the observed mass). In some models, the explosion energy has uncertainty, and \cite{Burrowsetal2024} give the energy in a range. In the line in Figure \ref{fig:Expenergy}, I take the upper energy in this range. In the range from a gravitational NS mass of $M_{\rm NS} = 1.361 M_\odot$, which results from a model of $M_{\rm ZAMS}=11M_\odot$, and to $M_{\rm NS} = 1.47 M_\odot$, which results from a model of $M_{\rm ZAMS}=15M_\odot$ I draw a thin-orange line. The reason is that the models of $M_{\rm ZAMS}=12.25M_\odot$ and $M_{\rm ZAMS}=14M_\odot$ do not explode in the simulations of \cite{Burrowsetal2024}.
 I did not present in the graphs the models of 
$M_{\rm ZAMS}=25M_\odot$ that ends with $M_{\rm NS} = 1.83 M_\odot$ and explosion energy of $e_{\rm exp} = 1.10 - 1.40 {\rm foe}$, and of 
$M_{\rm ZAMS}=60M_\odot$ that ends with $M_{\rm NS} = 1.59 M_\odot$ and explosion energy of 
$E_{\rm exp} = 0.54 - 0.72 \erg$, because the fraction of stars with $M_{\rm ZAMS} \ge 25 M_\odot$ is very small. I am here to account for the majority of CCSNe. 

The models in the simulations of \cite{Burrowsetal2024} that explode with energies of $E_{\rm exp} \gtrsim 0.4 \times 10^{51} \erg$ gain the energy over a timescale of $\gtrsim 4 \s$. \cite{Bolligetal2021} simulate one case where the explosion process lasts for about 7 seconds, ending with a NS mass of $M_{\rm NS,g}=1.675 M_\odot$  and an explosion energy of $E_{\rm exp}=0.94 \times 10^{51} \erg$. 

The simulations by \cite{Burrowsetal2024}, if explode and leave an NS, yield explosion energy of  $E_{\rm exp} \gtrsim 0.35 \times 10^{51} \erg$ for $M_{\rm NS,g} \gtrsim 1.5M_\odot$. The NS mass distribution shows that less than half of NSs fall in this mass limit.  Models with $E_{\rm exp} \gtrsim 0.65 \times 10^{51} \erg$ have $M_{\rm NS,g} \gtrsim 1.7M_\odot$. 
The NS mass distribution shows that only a minority of NSs obey this constraint. \cite{Mezzacappaetal2014}, for example, compared their results with eight observed CCSNe with explosion energies of $E_{\rm exp,obs} \gtrsim 0.8 \times 10^{51} \erg$. In the sample of CCSNe that \cite{Martinezetal2022} study about $75 \%$ of the CCSNe have $E_{\rm exp} \gtrsim 0.5 \times 10^{51} \erg$.   

In another recent paper, \cite{JankaKresse2024} conducted simulations of the neutrino mechanism and listed the explosion energy versus final (after cooling) NS mass. I placed their eight simulated cases with green circles in Figure \ref{fig:Expenergy}. Their results are similar to those of \cite{Burrowsetal2024} and strengthen the conclusion that only a small fraction of the simulated neutrino mechanism explosions yield the typical explosion energies of $E_{\rm exp} \simeq 10^{51} \erg$. 

In a recent study, \cite{Hiramatsuetal2025} analyzed Type IIn CCSNe and inferred higher explosion energies than those achievable with the neutrino mechanism for some of these CCSNe. 

My conclusion of this short comparison of simulations of \cite{Burrowsetal2024} and \cite{JankaKresse2024} with the observed mass distribution of NSs is that the explosion mechanism that they simulated might account for only a small fraction of observed CCNSe. The JJEM asserts that jets will take over and explode the star before neutrino heating expels core material.  
The simulations do show that neutrino heating is significant. In the JJEM, neutrino heating boosts the explosion but does not play the primary role in the explosion process \citep{Soker2022nu}.

% =========================================
\section{Seed perturbations by pre-explosion convection}
\label{sec:Convection}
% =========================================
% =========================
\subsection{Perturbations in the two explosion mechanisms}
\label{subsec:Perturbations}
% =========================
Both the neutrino mechanism and the JJEM require pre-collapse perturbations in the core. It seems that both mechanisms require perturbations of the same magnitude. In the neutrino mechanism, the need for perturbations and their magnitude have been determined by simulations. In contrast, in the JJEM, the needed perturbation magnitude in specific angular momentum is still a free parameter. I elaborate on these here. 

To facilitate explosion, (at least some) 3D simulations of the neutrino mechanism require pre-collapse core perturbations of $\simeq {\rm few}-10 \%$ (point 9 in Section \ref{sec:comparison}). These are relative perturbations in the density and/or turbulence velocity-to-sound speed ratio.   
The perturbations make the surface of the stalled shock non-spherical by increasing the post-shock convection and by non-spherical ram pressure with fluctuations of a few to 20 percent (e.g., \citealt{Varmaetal2023}).  
\cite{Couchetal2015} have in their 3D simulation a convective velocity amplitude of $v_{\rm conv} \simeq 500 \km \s^{-1}$ at a radius of $r \simeq 2000 \km $. \cite{Vartanyanetal2019} introduce by hand velocity perturbation of a few percent. 
Other introduce by hand smaller perturbations of only $0.1\%$ in density (e.g., \citealt{Nakamuraetal2024}). 

The JJEM defines the specific angular momentum parameter as the convective velocity times the radius
\begin{equation}
    j_{\rm conv} \equiv v_{\rm conv} r .
    \label{eq:Vconv}
\end{equation} 
The JJEM assumes that the seed perturbations are amplified by instabilities behind the stalled shock and above the NS. 
A centrifugally-supported accretion disk on the surface of the NS has a specific angular momentum of $j_{\rm d,NS} \simeq 2 \times 10^{16} \cm^2 \s^{-1}$. Simulations of the delayed-neutrino mechanism show that in 3D, the convective velocity amplitude is about $\times 3$ larger than in 1D according to the mixing length theory (e.g., \citealt{FieldsCouch2020}). For the larger convective velocity amplitude in 3D stellar models and the amplification of perturbations behind the stalled shock, \cite{ShishkinSoker2022} assumed that the requirement on the specific angular momentum parameter to allow intermittent accretion disks is $j_{\rm conv} \gtrsim 0.1-0.25 j_{\rm d,NS} \simeq 2 \times 10^{15} - 5 \times 10^{15} \cm^2 \s^{-1}$, where the convective velocity is the value from the mixing length theory in 1D simulations (for earlier studies of the angular momentum fluctuations in the JJEM see, e.g., \citealt{GilkisSoker2014, GilkisSoker2016, ShishkinSoker2021}).

In the simulation of \cite{Couchetal2015} mentioned above $j_{\rm conv} \simeq 10^{16} \cm^2 \s^{-1}$ at $r\simeq 2000 \km$. 
\cite{Mulleretal2016} and \cite{Mulleretal2017} have a convective  Mach number of $Mach \simeq 0.05-0.1$ at $r \simeq 4000 \km$, corresponding to a convective velocity of $v_{\rm conv} \simeq 300 \km \s^{-1}$, implying $j_{\rm conv} \simeq 10^{16} \cm^2 \s^{-1}$. 

The above discussion shows that the perturbations that the neutrino mechanism requires to facilitate explosion will also allow the formation of intermittent accretion disks in the JJEM. This is a subject for the not-so-near future as present hydrodynamical codes do not yet have the resolution to simulate the JJEM \citep{Soker2024Learning}. 

% =========================
\subsection{Pre-explosion stellar expansion}
\label{subsec:Expansion}
% =========================

The ejecta of many CCSNe interacts with a compact circumstellar material (CSM), e.g., the two recent examples of SN 2023ixf (e.g., \citealt{JacobsonGalanetal2023, Bostroemetal2024}) and SN 2024ggi (e.g., \citealt{XiangD2024, ZhangJ2024}). This suggests some pre-collapse activity that inflates the envelope by deposition of energy. The inflated envelope might trigger a binary interaction that powers an outburst in case of the presence of a close companion \cite{Soker2013B}, might form an extended envelope (e.g., \citealt{Dessartetal2017, Soker2024Effer, FullerTsuna2024}), or an extended wind-accelerated zone (e.g., \citealt{Moriyaetal2017}). 
The energy source should be related to the nuclear evolution of the core to `know' about the coming explosion. This can be the excitation of waves by the vigorous core convection (e.g., \citealt{QuataertShiode2012, RoMatzner2017, WuFuller2021}), or convection-power enhanced magnetic activity in the core (e.g., \citealt{CohenSoker2024}).  

The relevant point for this study is that the power of the waves is related to the convective power, where the ratio is some power of the convective Mach number (e.g., \citealt{ShiodeQuataert2014, WuFuller2022}). 
For the uncertainty in the exact fraction of the convective power that ends in waves propagating to the envelope, I take it as a small factor $\eta_c<0.1$ to read  
\begin{equation}
    L_{\rm wave} \simeq \eta_c L_{\rm conv} = 
    \eta_c 4 \pi \rho (r) r^2 v^3_{\rm conv} =  4 \pi \eta_c \frac{\rho}{r} j^3_{\rm conv}. 
    \label{eq:Lwave}
\end{equation}
where the different quantities are taken at the radius where $L_{\rm conv}$ has its maximum value. 
The core magnetic activity also depends on the convective power and the differential rotation in the core. Equation (\ref{eq:Lwave}) shows that the envelope energy deposition increases substantially as the specific angular momentum parameter (equation \ref{eq:Vconv}) increases.  

I raise the possibility that the same vigorous convective core motion that develops years to seconds before the core collapses and facilitates explosion, either in the neutrino mechanism or in the JJEM, also deposits energy to the envelope (by exciting waves or enhancing magnetic activity). The common source of pre-explosion envelope activity and seed perturbations that facilitate explosion deserves a deep study. 

% ================================================
\section{Summary}
\label{sec:Summary}
% ================================================

This paper compares the two alternative explosion mechanisms of CCSNe, mainly in light of new results from 2024 (Section \ref{sec:comparison}). I elaborated on two points. In Section \ref{sec:Energy} I examined two 2024 papers that present simulations of the neutrino mechanism. In Figure \ref{fig:Expenergy}, I present these studies' explosion energy versus the final NS mass. I compared the distribution of the NS from the simulations with that from observations. I concluded that to account for the distribution of explosion energies of observed CCNSe, the simulations produce NS masses that are higher than those observed. This conclusion strengthens the energetic problem of the neutrino mechanism, i.e., it cannot supply explosion energies of $E_{\rm exp} > 2 \times 10^{51} \erg$. Future simulations of the neutrino mechanism may result in higher explosion energies. 

The two mechanisms differ substantially in two observables. The first is the neutrino mechanism's prediction of a large population of `failed supernovae,' namely, stars that collapse into a black hole and fade with only a minor outburst. In the JJEM, even if most of the star collapses to a black hole, the outer layer of the star possesses significant angular momentum fluctuations to launch jittering jets that power an explosion, even a super-energetic one \citep{Gilkisetal2016}. In point 5 of Section \ref{sec:comparison}, I mentioned the recent claim for a `failed supernova,'  M31-2014-DS1, and based on the JJEM, I predicted that it would reappear within several years. The fading star might be a type II ILOT. The existence or absence of `failed supernovae' is under debate and should be decided in several years concerning the two `failed supernova' candidates, N6946-BH1 and M31-2014-DS1. 

The most significant difference between the two mechanisms (point 7 of Section \ref{sec:comparison}) is the JJEM's prediction that some, but not all, CCSNRs possess point-symmetrical morphologies. Identifying point-symmetrical morphologies in 12 CCSNRs, most in 2024 (Section \ref{sec:Introduction}), some with robust point-symmetrical morphologies, challenges the neutrino mechanism to the degree that it might rule out the neutrino mechanism as the primary explosion mechanism of CCSNe. 

The Crab Nebula explosion energy is low, $E_{\rm exp,Crab} \simeq 10^{50} \erg$ (e.g.,  \citealt{YangChevalier2015, BietenholzNugent2015}), and therefore, the modelers of the neutrino mechanism use it as an example of a CCSN for which the low explosion energies of their simulations apply. However, very recently \cite{ShishkinSoker2025Crab} identified point-symmetrical morphological features in the Crab Nebula that they attributed to jittering jets. Namely, even the Crab Nebula, a low-explosion-energy CCSN, exploded with jittering jets, i.e., the JJEM.

Putting together the success of neutrino mechanism simulations in achieving explosions of stellar models, even if with low energies, and the point-symmetric morphologies of CCSNRs, even in the Crab Nebula, I conclude that the neutrino mechanism is not the primary explosion process but that neutrino heating does play a role in the explosion process; neutrino heating boosts the jittering jets \citep{Soker2022nu}. The absence of a large population of `failed supernovae' supports this conclusion. Jets start to operate before the neutrino mechanism might have operated.

The demand for pre-explosion core seed perturbations is common to the neutrino mechanism and the JJEM (point 9 in Section \ref{sec:comparison}). These are density and/or velocity perturbations in the neutrino mechanisms and specific angular momentum perturbations in the JJEM. The pre-collapse core convection supplies these seed perturbations.
In section \ref{sec:Convection}, I deduced that the magnitude of required perturbations is about the same in both mechanisms. I further raised the possibility that the necessary vigorous core convection before explosion also triggers core activity that deposits energy to the envelope, which leads to pre-explosion activity that enhances mass loss. This deserves further study. 

I summarize that researchers of both explosion mechanisms have claimed success in 2024. As I see it, the success of the neutrino mechanism's studies is in showing that neutrino heating is non-negligible, and that of the JJEM is in establishing it as the primary explosion mechanism, boosted by neutrino heating.

% ===================================================
\section*{Acknowledgements}
% ===================================================
A grant from the Pazy Foundation supported this research.

%%%%%%%%%%%%%%%%%%%%%%%%%%%%%%%%%
% \bibliography{bib.bib}
%%%%%%%%%%%%%%%%%%%%%%%%%%%%%%%%%

\end{document}